\documentclass[
showpacs,
floatfix,
aps,
prl,
twocolumn,
superscriptaddress,
amssymb,
]{revtex4-1}

\usepackage{graphicx}
\usepackage{bm}
\usepackage{color}
\usepackage[normalem]{ulem}
\usepackage{amsmath}
\usepackage{cleveref}

\newcommand{\be}{\begin{equation}}
\newcommand{\ee}{\end{equation}}
\newcommand{\bea}{\begin{eqnarray}}
\newcommand{\eea}{\end{eqnarray}}

\def\wse2{WSe$_2$}

\newcommand{\rfig}[1]{Fig.\,\ref{#1}}

\begin{document}

\title{Odd- and even-denominator fractional quantum Hall states in monolayer \wse2}

\author{Qianhui Shi}
\affiliation{Department of Physics, Columbia University, New York, NY, USA}
\author{En-Min Shih}
\affiliation{Department of Physics, Columbia University, New York, NY, USA}
\author{Martin V. Gustafsson}
\affiliation{Department of Physics, Columbia University, New York, NY, USA} 
\affiliation{Raytheon BBN Technologies, Cambridge, MA 02138, USA}
\author{Daniel A. Rhodes}
\affiliation{Department of Mechanical Engineering, Columbia University, New York, NY, USA}
\author{Bumho Kim}
\affiliation{Department of Mechanical Engineering, Columbia University, New York, NY, USA}
\author{Kenji Watanabe}
\affiliation{National Institute for Materials Science, 1-1 Namiki, Tsukuba, Japan}
\author{Takashi Taniguchi}
\affiliation{National Institute for Materials Science, 1-1 Namiki, Tsukuba, Japan}
\author{Zlatko Papi\'c}
\affiliation{School of Physics and Astronomy, University of Leeds, Leeds LS2 9JT, UK}
\author{James Hone}
\affiliation{Department of Mechanical Engineering, Columbia University, New York, NY, USA}
\author{Cory R. Dean}
\affiliation{Department of Physics, Columbia University, New York, NY, USA}

\date{\today}

%\pacs{72.80.Vp, 73.22.Pr, 73.63.-b}

%\begin{abstract}

%\end{abstract}

\maketitle

\textbf{
Monolayer (ML) semiconducting transition-metal dichalcogenides (TMDs) represent a unique
class of two-dimensional (2D) electron systems. 
Their atomically thin structure facilitates gate-tunability just like graphene, but unlike the latter, TMDs have the advantage of a sizable band gap and strong spin-orbit coupling.
Measurements under large magnetic fields have revealed an unusual Landau level (LL) structure \cite{wang:2016,movva:2017,gustafsson:2018}, distinct from other 2D electron systems.
However, owing to limited sample quality and poor electrical contact, probing the lowest LLs has been challenging, and observation of electron correlations within the fractionally filled LL regime has not been possible.  
Here, through bulk electronic compressibility measurements, we investigate the LL structure of ML \wse2 in the extreme quantum limit, and observe fractional quantum Hall (FQH) states in the lowest three LLs.
The odd-denominator FQH sequences demonstrate a systematic evolution with the LL orbital index, which has not been observed in any other system but is consistent with generic theoretical expectations.
In addition, we observe an even-denominator state in the second LL that is expected to host non-Abelian statistics. 
Our results suggest that the 2D semiconductors can provide an experimental platform that closely resembles idealised theoretical models in the quantum Hall regime.}

The fractional quantum Hall (FQH) effect results from strong Coulomb interactions driving a collective state within a partially filled Landau level (LL) \cite{tsui:1982b}.  The associated quasiparticle excitations are fractionally charged anyons with unconventional exchange statistics, providing one of the few opportunities to experimentally study an electronic system with fractional topological order.
Despite more than three decades of efforts, several fundamental questions remain. 
While the FQH states in the lowest LL are relatively well understood, those in higher LLs are not;
the $N = 1$ LL ($N$ is the LL orbital index), in particular, manifests delicate competition between FQH states and electronic solid states, and the FQH hierarchy exhibits parameter-dependent variations that do not follow a universal rule \cite{kleinbaum:2015}.
Additionally, an even-denominator state appears at half filling of the $N=1$ LL \cite{willett:1987,falson:2015,ki:2014,zibrov:2017,li:2017b} that is presumed to host non-Abelian Majorana excitations \cite{moore:1991,levin:2007,lee:2007,son:2015}, however, their definitive experimental verification is still lacking.

%Further, the anomalous even-denominator state, observed at half filling of the $N=1$ LL \cite{willett:1987,falson:2015,zibrov:2017,li:2017b}, is presumed to host Majorana excitations that obey non-Abelian fractional statistics \cite{moore:1991}. 
%While theoretical works have proposed a number of trial states to describe this fraction \cite{moore:1991,levin:2007,lee:2007,son:2015}, their definitive experimental verification is still lacking.

%due to complications of sample-dependent effects, such as finite thickness and multiple degenerate degrees of freedom.
%[I think you should also introduce the predicted FQH sequence / hierarchy in front, rather than at the end...]

In conventional heterostructures such as GaAs and ZnO \cite{tsui:1982b,willett:1987,falson:2015}, complications due to finite thickness effects are always present.
Moreover, the FQH states are generally fragile, with most states appearing only in the highest mobility devices at sub-Kelvin temperatures. In ML graphene, by contrast, wide gate tunability and weak screening enable observation of the FQH states at an order of magnitude higher temperatures. However, the unique features of graphene also create new challenges.  For example, the orbital wavefunction in the $N = 1$ LL of ML graphene is an equal mixture of the non-relativistic $|0\rangle$ and $|1\rangle$ harmonic oscillator wavefunctions \cite{goerbig:2011}, which does not favor formation of the even-denominator states.    Additionally, the four-fold degeneracy in monolayer graphene, due to the spin and valley degrees of freedom, further complicates the LL structure and may or may-not modify the FQH spectrum, depending sensitively on the device details \cite{dean:2011,feldman:2012,zibrov:2018}.    Finally, the lack of a robust band gap in graphene limits the ability to engineer electrostatically-tunable mesoscopic structures, such as quantum point contacts and interferometers - important tools for unveiling the potentially exotic properties of various FQH states \cite{chamon:1997}.

%%%%%%%%%%%%%%%%%%%%%%%%%%%%%%%%%%%%%%%%%%%%%%%%%%%%%%%%%%
\begin{figure*}
\centering
\includegraphics[width=1\linewidth]{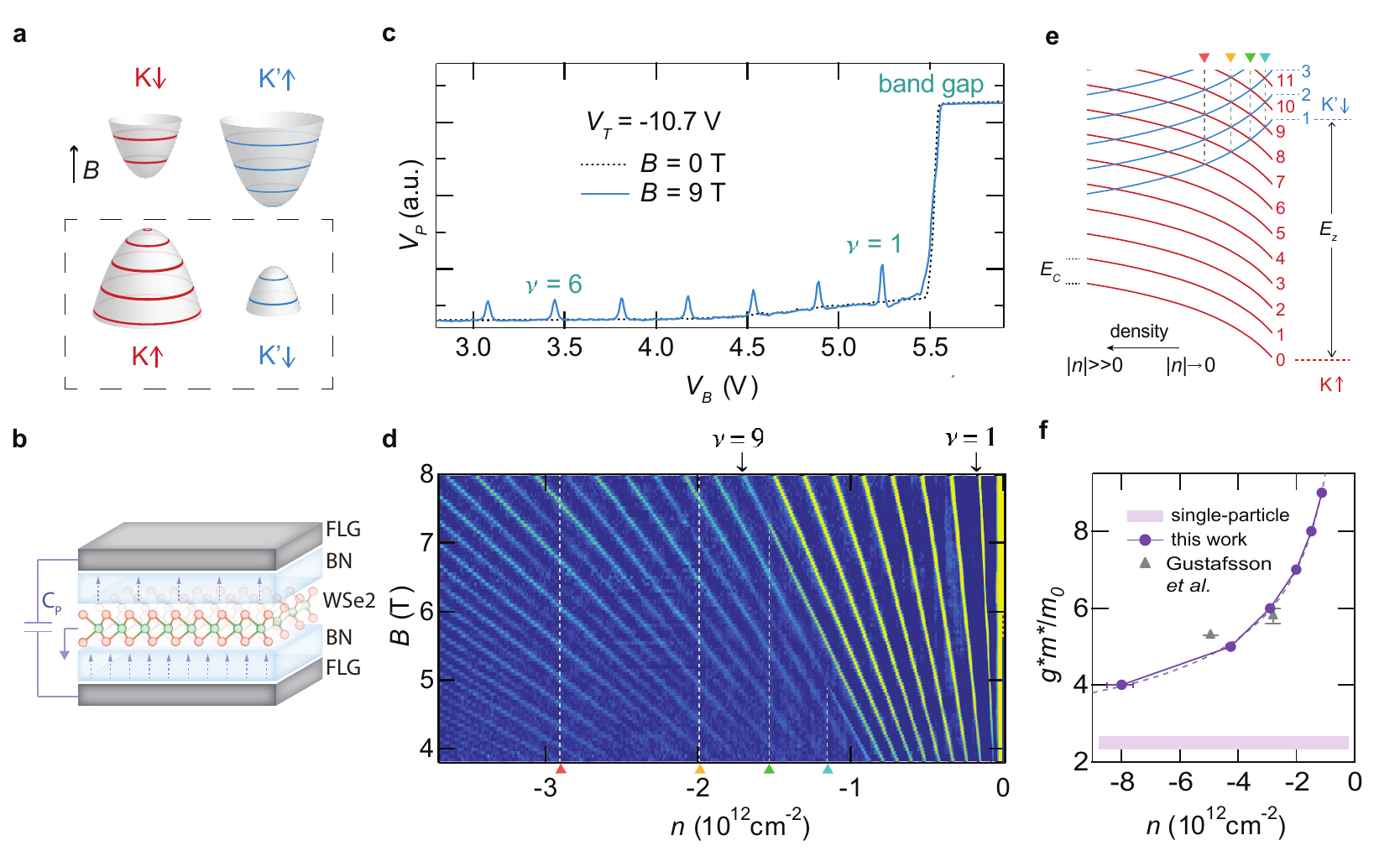}
\vspace{-0.15 in}
\caption{\small{
{\bf{Measurement scheme and Landau level structure.}}
(a) Cartoon of the low-energy band structure subject to a perpendicular magnetic field. The dashed line box highlights the valence band which is relevant in our study.
(b) Cartoon of the stack which illustrates the definition of penetration capacitance. 
ML \wse2 flakes encapsulated between two hexagonal boron nitride (hBN) layers and few-layer graphite (FLG) flakes serve as top and bottom gates.
(c) Penetration signal as a function of the back gate voltage $V_B$ at 0 T (dotted) and 9 T (solid), with top gate voltage $V_T = - 10.7$ V.
(d) Penetration signal versus carrier density and the magnetic field measured at $T =$ 1.6 K.
The vertical dashed lines mark the densities where the contrast between strong and weak gaps is the most prominent.
(e) Schematic illustration of the evolution of Landau level diagram as carrier density is varied.
Red and blue lines stands for spin up and spin down branches respectively, and the numbers on the right mark the LL orbital indices.
The dashed lines illustrate the same densities as those in \rfig{fig1}(d), where level (anti-)crossings between LLs leads to the strongest variation of gaps at neighboring integer filling factors.
(f) Magnetic susceptibility $\chi = E_z/E_c = g^* m^*/m_0$ as a function of carrier density.
The thick horizontal line marks the value from bare $g$ factor and effective mass from a single-particle picture.
The triangles mark the same quantity reported in literature.
}
}
\label{fig1}
\vspace{-0.15 in}
\end{figure*}
%%%%%%%%%%%%%%%%%%%%%%%%%%%%%%%%%%%%%%%%%%%%%%%%%%%%%%%%%%%%%%

Semiconducting TMDs such as ML \wse2 provide a potential opportunity to bridge the gap between conventional heterostructures and graphene. 
%Belonging to the class of van der Waals materials, the TMDs can be mechanically exfoliated to the ML limit. However, unlike graphene they can possess a large band-gap and strong spin-orbit coupling.  
The spin and valley degrees of freedom are locked [see \rfig{fig1}(a)] \cite{xiao:2012}, which, together with an unusually large spin-splitting \cite{gustafsson:2018}  yields a LL spectrum where the low energy levels are fully spin- and valley-polarized, effectively eliminating the degenerate degrees of freedom \cite{gustafsson:2018}. 
%Similar to graphene, the $N = 1$ LL wavefunction in \wse2 is formally a mixture of $|0\rangle$ and $|1\rangle$ \cite{li:2013,rose:2013}, but because of the large band gap, the weight of the $|0\rangle$ component is only a few percent and so the $|1\rangle$ component dominates~\cite{rose:2013}. 
Although the $N = 1$ LL orbital wavefunction in \wse2 is formally also a mixture of $|0\rangle$ and $|1\rangle$ \cite{li:2013,rose:2013} similar to graphene, the large band gap suppresses the weight of the $|0\rangle$ component to only a few percent and the $|1\rangle$ component dominates~\cite{rose:2013}.

Although the LL structure of semiconducting TMDs has been experimentally probed \cite{wang:2016,gustafsson:2018,larentis:2018,pisoni:2019}, observation of the FQH states has been elusive. This is largely due to a combination of limited material quality and poor electrical contact, both of which conspire to make the low density regime, where the FQH states may be observed, experimentally inaccessible \cite{movva:2017,gustafsson:2018}. Here we address both issues simultaneously by performing capacitance measurements on high-quality \wse2. The samples were grown by a previously reported, self-flux method, which yields a density of atomic defects below $10^{11}/\mathrm{cm}^2$~\cite{edelberg:2019}.
The reduced sensitivity of the capacitance measurement to both large contact resistance and disorder-induced-localization allows us to probe the electronic compressibility in the valence band all the way to the band edge.

%%%%%%%%%%%%%%%%%%%%%%%%%%%%%%%%%%%%%%%%%%%%%%%%%%
\begin{figure*}
\begin{center}
\includegraphics{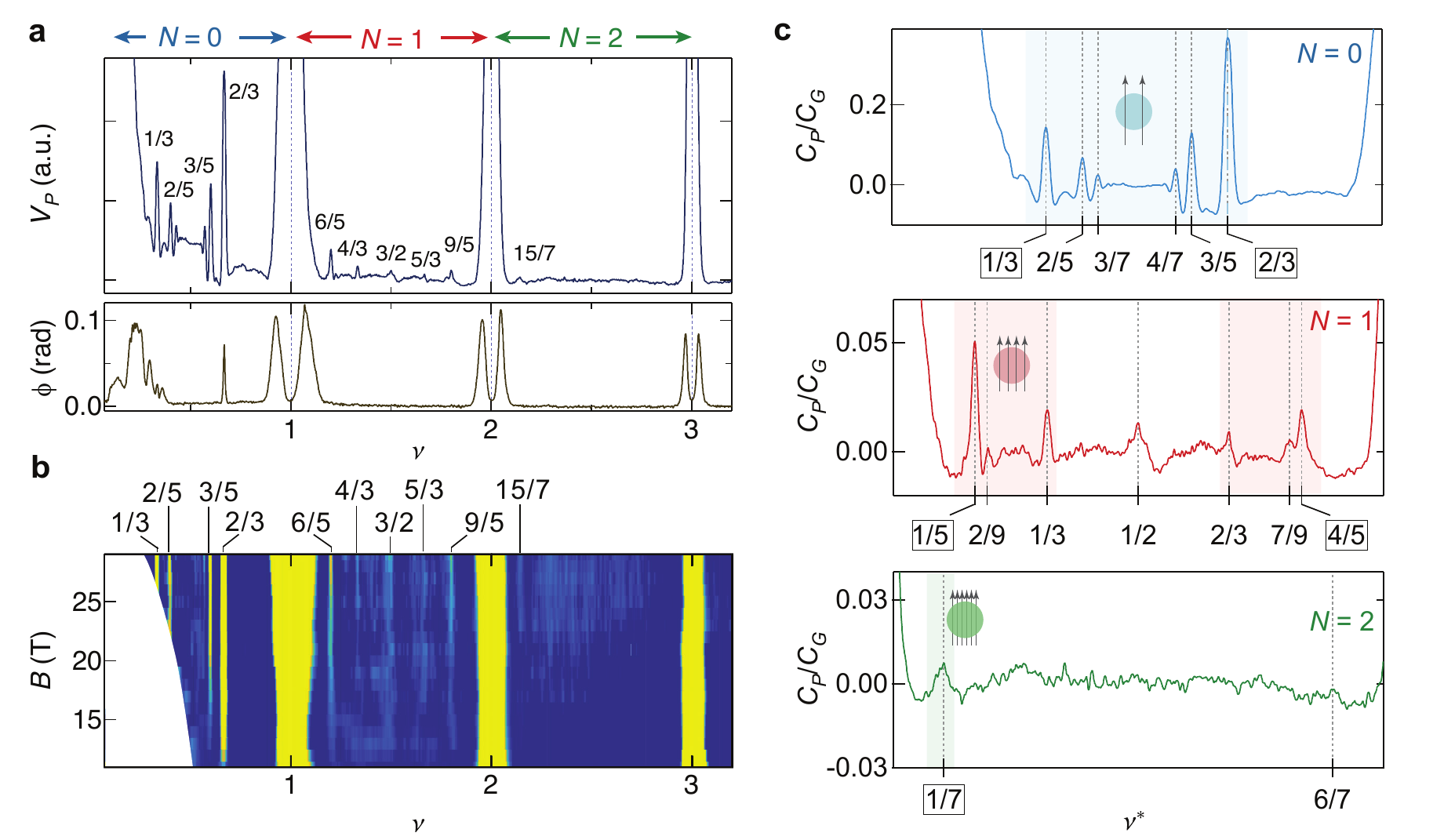}
\vspace{-0.15 in}
\end{center}
\caption{\small{
{\bf{Fractional quantum Hall states.}}
(a) Penetration signal (top) and the phase (bottom) versus filling factor $\nu$.
(b) Penetration signal versus filling factor $\nu$ and magnetic field $B$.
(c) Penetration capacitance normalized by the geometry capacitance between top and bottom gates, $C_P/C_G$, plotted versus the partial filling factor $\nu^*$, for $N = 0$ (top), $N = 1$ (middle) and $N=2$ (bottom) at $B = $ 29 T and $T$ = 0.3 K. 
Filling factors where incompressible states appear are marked with vertical dotted lines, with the filling factor of the most strong states (and their conjugates) for each LL boxed.
Cartoon illustrates the most prominent composite fermion construction for each LL, with the number of arrows illustrating the number of flux quanta attachment to each electron.
}}
\label{fig2}
\vspace{-0.15 in}
\end{figure*}
%%%%%%%%%%%%%%%%%%%%%%%%%%%%%%%%%%%%%%%%%%%%%%%%%%%%%%%%%%

\rfig{fig1}(b) shows a schematic of the device geometry and capacitance measurement scheme (see Methods and Supplementary section 1 for device details including contact configuration). The capacitance measurement closely follows similar techniques that have been applied to both GaAs and graphene in the quantum Hall regime~\citep{eisenstein:1994,zibrov:2017,zibrov:2018} (see Methods and Supplementary Fig.2).  
In brief, we apply a small ac signal, $V_{AC}$, to one of the gates, and measure the induced ac voltage, $V_P$, on the other, while the ML \wse2 is grouded.  In the absence of dissipation, $V_P$ is, to a good approximation,  proportional to the penetration capacitance, $C_P$, which itself is proportional to the inverse bulk electronic compressibility, $n^2 d\mu/dn$, where $n$ is carrier density and $\mu$ is the chemical potential of the \wse2 (see Supplementary section 2).  A small-valued $V_P$ therefore indicates that the \wse2 layer is compressible, whereas a large $V_P$ indicates that it is incompressible, \textit{i.e.} gapped.  
%In a typical measurement, we fix the top gate voltage $V_T$ and tune the density by varying the DC bias applied to the bottom gate, $V_B$ .

\rfig{fig1}(c) shows a plot of $V_P$ versus back gate voltage $V_B$, measured at both zero and finite magnetic field at $T = 1.6$ K.  
At large $V_B$, the Fermi level of ML \wse2 is in the band gap and the channel is insulating, yielding a plateau in $V_P$ that corresponds to the maximum measurable penetration signal. Under zero magnetic field (dotted line), the signal drops sharply as $V_B$ falls below $5.5$ V, indicating that the fermi level of ML \wse2 enters the valence band. The channel is then compressible with a constant finite density of states, resulting in a small penetration signal.
At $B = 9$ T (solid line), we observe the same general trend, but with multiple new peaks corresponding to the incompressible cyclotron gaps between the LLs. 
The peaks appear at integer filling factors $\nu = n h/eB$, where $n$ is the hole carrier density determined from the geometric capacitance between the channel and the two gates, $h$ is Planck's constant, and $e$ is the electron charge. Observation of a peak at all integer fillings indicates a fully lifted LL degeneracy.

In \rfig{fig1}(d) we plot the penetration signal versus the magnetic field $B$ and carrier density $n$ at $T = 1.6$ K. 
The LL gaps demonstrate variations in the hierarchy as density is varied, switching between odd and even integer $\nu$ dominated. 
This is consistent with previous reports in TMDs \cite{movva:2017,gustafsson:2018}, and results from level (anti-)crossings due to the exchange-enhanced and density-dependent spin-splitting energy between the two sets of LLs, $E_z = 2g^*\mu_B B$  ($g^*$ is the effective $g$-factor, $\mu_B = e\hbar/2m_0$ is the Bohr magneton), as sketched in \rfig{fig1}(e).
%\qs{This is consistent with previous reports in TMDs, and results from level crossings as the effective Zeeman splitting increases at lower carrier density due to the enhancement of electron-electron interactions [\rfig{fig1}(e)] \cite{movva:2017,gustafsson:2018}.} 
By tracking the density where the level crossings occur (see Supplementary section 3), we determine the spin-susceptibility $\chi = E_z/E_c = g^* m^*/m_0$  as a function of density [\rfig{fig1}(f)], where $E_C$ is the cyclotron energy and $m^*$ is the effective mass of hole carriers.
The capability to extend to the low density regime reveals that $E_Z$/$E_{C}$ can be as large as 9, \textit{i.e.} 60\% larger than previously identified and more than three times larger than the single-particle value, further attesting to the strong electron interaction effects in this system.
Importantly, these results confirm a simple LL structure with fully lifted degeneracy for the lowest LLs -- the regime where FQH states are expected -- which we discuss below.

%%%%%%%%%%%%%%%%%%%%%%%%%%%%%%%%%%%%%%%%%%%%
\begin{figure*}
\begin{center}
\includegraphics{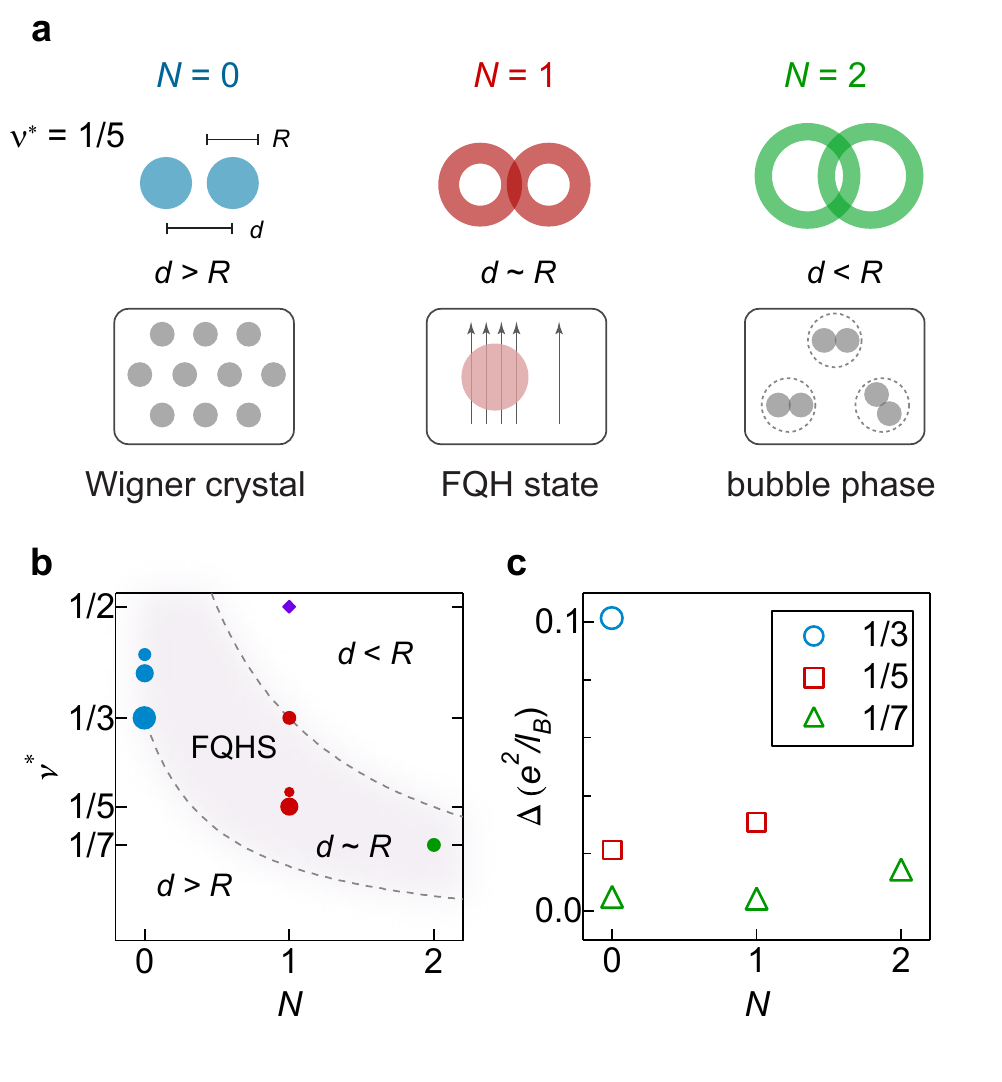}
\vspace{-0.15 in}
\end{center}
\caption{\small{
{\bf{Evolution of fractional quantum Hall states with Landau level orbital index.}}
(a) Cartoon illustrating the evolution of robustness of FQH states at different $N$ for fixed filling factor $\nu^*= 1/5$.
FQH states are the favorable ground states when the two lengths scales $d$ and $R$ are comparable.
The disks and rings are drawn to qualitatively represent the wavefunctions in different LLs (in the symmetric gauge with angular momentum $m = 0$).
A ring shape is a good representative of the high-LL wavefunctions as the interaction potential assumes a plateau-like behaviour for $r < R$ \cite{goerbig:2003}.
(b) Schematic phase diagram for FQH states.
Odd- (even-)denominator states are marked by circles (diamonds).
The area of the circle is proportional to the square root of the charge gap (integrated from $\partial\mu/\partial n$).
The dashed lines are drawn at $R = \alpha d$ as suggested by \cite{goerbig:2003,goerbig:2004} and the factors $\alpha$ are chosen to bound the observed odd-denominator states.
(c) Numerically calculated charge gap for different filling fractions as a function of the LL index $N$.
}}
\label{fig3}
\vspace{-0.15 in}
\end{figure*}
%%%%%%%%%%%%%%%%%%%%%%%%%%%%%%%%%%%%%%%%%%%%%%%%%%

Now we focus on the lowest LLs under very large magnetic fields. 
In the top panel of \rfig{fig2}(a), the penetration signal $V_P$ measured at $B =$ 29 T and $T = 0.3$ K is plotted for filling factor range $0 < \nu < 3$.  The top axis labels the corresponding orbital index for each filling range (see also Supplementary section 4 and 5).
In addition to large $V_P$ at integer $\nu$, we observe peaks in $V_P$ at fractional $\nu$ - which signals the chemical potential discontinuity due to the formation of FQH states - in all three LLs, including one at seventh fractional filling in the $N = 2$ LL, which has not been observed before in any other system.
%In addition to the IQH states, we observe FQH states - seen as peaks in $V_P$ at fractional $\nu$ - in all three LLs,  including one at seventh fractional filling in the $N = 2$ LL, which has not been observed before in any other system.
For all states except $\nu = 1/3$ and $2/3$, the phase due to the dissipative contribution shows small and featureless response (Fig. 2a, bottom panel). 
The magnetic field dependence of the penetration capacitance is shown in \rfig{fig2}(b) from $B = 11$ T to 29 T. The features that we have labelled evolve along vertical trajectories (\textit{i.e.} fixed filling fraction) and generally become better developed with increasing field, consistent with our interpretation that these represent the FQH states. 

The FQH sequence shows a different hierarchy in each LL.
We denote the fractional part of the filling factor as $\nu^* = \nu - N$ and plot the penetration capacitance normalized by the geometry capacitance, $C_P/C_G$, versus  $\nu^*$, for the three LLs with orbital index $N = 0$, 1 and 2 in \rfig{fig2}(c).
The state that manifests the largest variation in $C_P$ (or its particle-hole conjugate at $1-\nu^*$) appear at $\nu^* = $ 1/3 in $N=0$, 1/5 in $N=1$, and the 1/7  in $N=2$ LL.
This observation can be summarized  by the simple rule that, in each LL, the most robust FQH state is at $\nu^*=1/(2N+3$).

In addition to the strongest individual state in each LL, the overall sequence also shows a systematic trend. In the $N=0$ LL the FQH hierarchy follows the standard sequence that is observed in other 2D systems, with fractional  states appearing at  $\nu^* = n/(2n+1)$ where $n$ takes integer values, and with the relative strength of each state diminishing with increasing $n$.  This hierarchy is well described by the composite fermion (CF) model in which strong electron-electron interactions are renormalized by attaching an even number of flux quanta to each electron (in this case 2 flux per electron), resulting in weakly-interacting CFs in reduced magnetic fields \cite{jain:book}. The 2-flux FQH sequence, such as observed here, resembles an integer quantum Hall sequence but centered around filling fraction $\nu=1/2$. In the  $N=1$ LL, the FQH hierarchy changes. Here the $\nu^* = 1/5$ and $4/5$ states are the strongest, with smaller peaks also  appearing at $2/9$ and $7/9$.  This hierarchy is consistent with a 4-flux sequence [$\nu^* = n/(4n+1)$)] centered around 1/4 and 3/4 filling.  Finally, for $N = 2$, the only clearly resolved FQH state appears at $\nu = 1/7$, which is the dominant fraction expected in a  6-flux CF sequence [$\nu^*=n/(6n+1)$]. This trend can also be summarized with a simple rule, namely, the most robust CF sequence observed in each LL corresponds to a CF $2p$-flux attachment series where $p=N+1$.

The systematic evolution of FQH sequence with LL orbital index results from the change of orbital wavefuntion and interaction potential.
Besides FQH states, strong interactions within a LL can drive competing ground states such as the electron solid phases \cite{goerbig:2003,goerbig:2004}.
The LL-dependence that we observe is qualitatively consistent with a previously identified scaling argument~\cite{goerbig:2004} that considered this competition. 
%The competition depends on the interaction potential which changes with LL index. 
%could be understood as resulting from its competitions with electron solid phases.
%
%The LL-dependence that we observe is qualitatively consistent with a previously identified scaling argument~\cite{goerbig:2004} that considered this competition. 
In this view, FQH states are favored when the inter-particle spacing, $d\sim l_B/\sqrt{\nu}$, is comparable to the interaction length scale, $R\sim 2R_C = 2\sqrt{2N+1} l_B$ ($R_C$ is the cyclotron radius and $l_{B}$ is the magnetic length). 
The FQH states give way to a Wigner solid when $d>R$, or charge density wave states (stripe and bubble phases) when $d<R$. An example is illustrated schematically in \rfig{fig3}(a) for the case of $\nu^*=1/5$, which according to this argument would favour a FQH ground state in the $N=1$ LL.  

Fig. 3b shows a schematic phase diagram  summarizing the FQH states observed in our sample. Consistent with the scaling argument, odd-denominator states are observed within a band corresponding to $d \sim R$. Likewise, as the orbital index increases, the range of filling factors over which the FQH states are resolved decreases. 
We note that we do not identify unambiguous signatures of the solid phases. It is possible that the large response in the dissipative signal in \rfig{fig2}(b) is due to the insulating Wigner crystal states, the boundaries of which move towards integer filling with increasing $N$.
%but we cannot rule out single-particle localization states as the origin of this response, due to LL-dependent disorder effects.

The LL dependence we observe also shows excellent agreement with exact diagonalization calculations of the quasiparticle-quasihole excitation energies.
In \rfig{fig3}(c) we plot the charge gap $\Delta$ (\textit{i.e.}, the sum of energies required to create a single quasihole and a single quasielectron) for filling factor $\nu^* = 1/3$, $1/5$ and $1/7$ as a function of $N$. 
It can be observed that the most robust FQH states for $N = 0$, 1, 2 are 1/3, 1/5, 1/7, respectively, consistent with our experimental data.
\rfig{fig3}(c) only shows the gaps for states that allow a reliable extrapolation to the thermodynamic limit, which coincide with the states that have a large overlap with the Laughlin wavefunction (see Supplementary section 8).

Finally, we observe an even-denominator state at $\nu = 3/2$, \textit{i.e.}, half filling of the $N = 1$ LL, as shown in the middle panel of \rfig{fig2}(c) and as the diamond mark in \rfig{fig3}(b).
The exclusive appearance of an even-denominator FQH state in the $N = 1$ LL \cite{willett:1987,falson:2015,ki:2014,zibrov:2017,li:2017b} is consistent with a Pfaffian \cite{moore:1991}, anti-Pfaffian \cite{levin:2007,lee:2007} or PH-Pfaffian wavefunction \cite{son:2015}, which is expected to host non-Abelian statistics (see Supplementary section 9 for more discussions).
Unlike other systems that have complications from multiple spin/valley degrees of freedom making it difficult to disambiguate among other candidate even-denominator states such as the Halperin (331) state ~\cite{halperin:1983,zibrov:2018}, the ML \wse2 is isospin polarized and only one set of LLs is present at low carrier densities, so that the possibility of a two-component state can be ruled out. Single-component even-denominator FQH states have only been observed in a handful of high mobility systems, including GaAs~\cite{willett:1987,pan:1999b}, ZnO~\cite{falson:2015}, and graphene~\cite{ki:2014,zibrov:2017,li:2017b}. 
This makes our observation especially surprising since prior transport studies suggest that the \wse2 mobility is limited to 30,000 Vs/cm$^{-2}$ \cite{pistunova:2019}.   One way the negative effect of disorder is addressed in our experiment is by applying large magnetic fields where the Coulomb interaction is maximized (At $B=29$~T, we observe a thermodynamic gap of about 1.6 K at $\nu = 3/2$).
We also note that the simple LL structure and wavefunctions in ML \wse2 allows the even-denominator states to be strengthened at arbitrarily large magnetic fields, which was not possible in other systems due to concomitant complications such as population of higher subbands in finite-thickness heterostructures or severe modification of orbital wavefunction in graphene \cite{ki:2014,zibrov:2017,li:2017b}. Additionally, numerical calculations predict that a slight mixing of $|0\rangle$ in the $|1\rangle$ orbital wavefunctions, as is the case in the $N=1$ LL of \wse2, optimizes the non-Abelian paired state ~\cite{apalkov:2011,papic:2011}.

While the overall FQH evolution we observe is consistent with theoretical expectations, such  a trend has not been experimentally identified in previously studied systems.  
For example, in most cases the 1/3 and 2/3 states remain the most prominent odd-denominator FQH in the  $N = 1$ LL (\textit{i.e.}, stronger than 1/5) \cite{pan:1999b,falson:2015,zibrov:2017}, while opposite behaviour is demonstrated in the two spin branches \cite{kleinbaum:2015}.
%while the 1/5 is comparatively weak or absent (opposite to our observations) \cite{pan:1999b,falson:2015,zibrov:2017}. 
In the $N=2$ LL where we see the 1/7 state, the only FQH state previously reported in any system appeared at $\nu^* = 1/5$ \cite{gervais:2004,zeng:2019}. 
The deviation from theoretical models in previous systems is presumed to result from a variety of sample-specific properties such as finite thickness effects \cite{kleinbaum:2015,peterson:2008}, multi-component degrees of freedom \cite{dean:2011,feldman:2012,zibrov:2018},  or more complex admixture wavefunctions \cite{goerbig:2011}.
The systematic evolution of FQH hierarchy observed in ML \wse2, by comparison, suggests that the semiconducting TMDs are a nearly ideal quantum Hall system in which experiments can be directly compared to simplified theoretical models (see Supplementary section 6).
%We note that, the LL mixing, characterized by the ratio between Coulomb interaction and cyclotron energy, is large in ML \wse2.
The LL mixing parameter, characterised by the ratio between the Coulomb interaction energy scale and LL spacing, $\kappa =\frac{e^2}{\epsilon l_B}/\hbar \omega_c \propto B^{-1/2}$, is large in ML \wse2 -- $\kappa \approx 7$ at $B = 29$ T.
Nevertheless, while LL mixing may be responsible for the particle-hole asymmetry in the FQH gap values in our sample, theoretical works have suggested that it does not significantly modify the FQH sequence \cite{sreejith:2017} (see Supplementary section 7).
%The LL mixing factor, characterized by the ratio between Coulomb interaction and cyclotron energy, is relatively large, may be responsible for the particle-hole asymmetry in the gap values, but 

To conclude, our compressibility measurements on ML \wse2 reveal FQH states including those that have either only been found in the highest-mobility systems or never been observed before.
These observations attest to the advancement of the intrinsic quality of the 2D semiconductors and establish them as a platform that is unique and sufficiently clean for further exploration of correlated electronic states.

\section{Methods}

\subsection{Device fabrication}

The device was built from exfoliated van der Waals materials using a dry transfer method in the following steps.  
First, hBN and graphite were picked up layer by layer and released on a silicon substrate, to serve as the bottom dielectric and gate. 
Second, Pt electrodes were evaporated onto the hBN and cleaned by an AFM tip in the contact mode. 
Third, hBN, graphite, hBN and monolayer WSe2 were picked up layer by layer and released onto the bottom hBN with pre-patterned Pt electrodes. 
All transfer was assisted by Polypropylene Carbonate on polydimethylsiloxane on a glass slide, with picking up temperature about 40 C and releasing temperature about 120 C.
Finally, the top graphite was etched to define the top gate so that the cross section between the top and bottom gates covers uniformly ML \wse2.
Fig.1 in the supplementary material shows the optical micrograph image of the device.

\subsection{Measurements}

Carrier density in ML \wse2 is controlled by DC gate voltages $V_T$ and $V_B$, $n = ( C_t V_T + C_b V_B)/e$ where $C_t$ ($C_b$) is the geometry capacitance between \wse2 and the top (bottom) gate.
The top gate is held at a constant high negative voltage in order to tune the \wse2 in contact with Pt to high density, allowing sufficient electrical contact to charge the \wse2 flake, and the bottom gate is swept to tune density.
For our ML \wse2, the measured capacitance does not appear to show any dependence on the displacement field $D \sim C_t V_T - C_b V_B$.
A ac excitation $\delta V_{ex}$ is superposed on $V_B$, and the penetration signal is detected from the top gate with the help of a FHX35X high electron mobility transistor (HEMT). 
The HEMT is biased with a DC current $I_H$ and its working point is set by the gate voltage $V_H$, both controlled independently from $V_T$ and $V_B$.
The ac voltage drop across the HEMT $\delta V_{out}$ is measured using lock-in amplifiers. 
Presented data are measured at 13.353 kHz using a 4 mV excitation.
The capacitance measurement scheme is illustrated in Supplementary Fig.2.

\section{Data availability}
Relevant data are available from the corresponding author upon reasonable request.

%\bibliographystyle{naturemag_noURL}
%\bibliographystyle{apsrev4-1}
%\bibliography{biba}

\section{Acknowledgement}
We thank Mark Goerbig, Andor Kormanyos, Fan Zhang for discussion and William Coniglio, Bobby Pullum for help with experiments. 
This research is primarily supported by US Department of Energy (DE-SC0016703). 
Synthesis of \wse2 (D.R. and B.K.) was supported by the Center for Precision Assembly of Superstratic and Superatomic Solids, a Materials Science and Engineering Research Center (MRSEC), through NSF grant DMR-1420634. 
A portion of this work was performed at the National High Magnetic Field Laboratory, which is supported by National Science Foundation Cooperative Agreement No. DMR-1157490 and the State of Florida. 
Z.P. acknowledges support by EPSRC grant EP/R020612/1.
K.W. and T.T. acknowledge support from the Elemental Strategy Initiative conducted by the MEXT, Japan and the CREST (JPMJCR15F3), JST.
%%%%%%%%%%%%%%%%%%%%%%%%%%%%%%%%%%%%%%%%

\section{Author Contributions}
Q.S. fabricated the device with the help of E.S.. M.V.G. contributed to the development of the measurement setup. Q.S. performed the measurements and analyzed the data. Z.P. performed the numerical calculations. D.A.R. and B.K. grew the \wse2 crystals. K.W. and T.T. grew the hBN crystals. J.H. and C.R.D. advised on the experiments. The manuscript was written with input from all authors.

\section{Additional information}

Supplementary information is available in the online version of the paper.

%%%%%%%%%%%%%%%%%%%%%%%%%%%%%%%%%%%%%%%%


\end{document}